\documentclass[pra,twocolumn]{revtex4}
\usepackage{graphicx}

\begin{document}
\title{Nonlinear optical properties of electromagnetically-induced-transparency medium  interacting with two quantized fields}
\author{Le-Man Kuang$^{1,2}$, Guang-Hong Chen$^{2}$,  and Yong-Shi Wu$^{2}$}
\address{$^{1}$Department of Physics, Hunan Normal University, Changsha 410081, People's Republic of China\\
$^{2}$Department of Physics, University of Utah,
Salt Lake City, UT 84112, USA}
%\date{\today }
%\maketitle
\begin{abstract}
We study  linear and nonlinear optical properties of
electromagnetically induced transparency (EIT) medium  interacting
with two quantized laser fields for adiabatic EIT case. We show
that EIT medium exhibits normal dispersion. Kerr and higher order
nonlinear refractive-index coefficients are also calculated in a
completely analytical form. It is indicated that EIT medium
exhibits giant resonantly enhanced nonlinearities. We discuss the
response of the EIT medium to nonclassical light fields and find
that the polarization vanishes when the probe laser is initially
in a nonclassical state of no single-photon coherence.

\noindent PACS numbers: 32.80.-t, 42.50.Gy, 42.65.-k
\end{abstract}
\maketitle

%\begin{multicols}{2}

\section{Introduction}
In the last ten years, much attention has been paid to
understandings and applications of systems exhibiting
electromagnetically induced  transparency (EIT) \cite{har}. EIT is
a powerful technique that can  be used to make an otherwise
absorbing medium transparent to a probe laser on resonance with a
certain atomic transition by applying a coupling laser under the
circumstances, while retaining the large and desirable nonlinear
optical properties associated with the resonant response of the
medium. Preparation of matter in such an EIT state would provide
us with a new type of optical material of interest both its own
right, and in many applications to fundamental and applied
physics. There has been a large number of theoretical
contributions in nonlinear optics using EIT
\cite{har,aga,ari,scu,fie}. The essence of EIT nonlinear optics is
to create strongly-coupled Raman coherence for a three-level
system and to control optical responses of the system. EIT has
been observed in several different experimental
 configurations \cite{fie,hak,gea}. Recently, considerable attention
has been drawn to slow group velocities of light and high
nonlinearities in the conditions of EIT. Extremely slow group
velocities  of light pulses have been observed in Bose-Einstein
condensate of  ultracold sodium atoms \cite{hau,ino}, in an
optically dense hot rubidium gas \cite{kas}, in rubidium vapor
with slow ground state relaxation \cite{bud}, and in crystals
doped by rare-earth ions \cite{tur}. It has been shown that the
condition of ultrslow light propagation leads to photon switching
at an energy cost of one photon per event \cite{hy} and to
efficient nonlinear processes at energies of a few photons  per
atomic cross section \cite{hh}. A giant cross-Kerr nonlinearity in
EIT was suggested by Schmidt and Imamo\v{g}lu \cite{sch}, and has
been indirectly measured in the experiment \cite{hau}. More
recently, two groups \cite{liu,phi} have independently  realized
light storage  in atomic mediums by using EIT technique.

Theoretically  there are two formalisms to treat  EIT.
One is adiabatic EIT \cite{har}
in which both  probe and coupling resonant lasers are adiabatically applied.
 After the system reaches a steady state, EIT occurs for arbitrary
intensities of the probe and coupling lasers. The other is
transient-state EIT \cite{scu}. In this case, resonant probe and
coupling lasers are simultaneously applied. EIT happens only when
the intensity of the coupling laser is much larger than that of
the probe.

Conventionally, both  coupling and probe lasers were treated as
classical, external fields. A disadvantage of the external field
approach is that it can not deal with atom-photon and
photon-photon quantum entanglement which is of importance not only
because of the fundamental physics involved, but also for their
potential technological applications such as quantum computation
and quantum communication  \cite{nil}. Recent studies on EIT have
indicated that the possibility to coherently control the
propagations of quantum probe light pulses in atomic media. This
opens up  interesting applications such as quantum state memories,
generation of squeezed  and  entangled atomic states, quantum
information processing, and as narrow-band sources of nonclassical
radiation. In particular,   with a quantum treatment of the probe
laser, Fleischhauer and Lukin \cite{fle} have recently been able
to predict the formation of  dark-state polaritons in the
propagation of light pulses through quantum entanglement of atomic
and probe-photon states. This has been confirmed  in the latest
light storage experiments \cite{liu,phi}. This lesson teaches us
that the quantum description of laser is more fundamental than the
classical one, having advantages in uncovering new effects
involving quantum nature of photons.

In a previous paper \cite{kua}, We have developed a fully quantum
treatment of EIT in a vapor of three-level $\Lambda$-type atoms.
Both the probe and coupling lasers with arbitrary intensities are
quantized, and treated on the same footing. The purpose of  the
present paper is to study the linear and nonlinear optical
properties of an adiabatic EIT system interacting with two
quantized fields and investigate the response of the EIT medium to
nonclassical light fields.

This paper is organized as follows. In Sec. II, we set up our
model and give an adiabatic solution of the full Hamiltonian. In
Sec. III and IV, we investigate the linear and nonlinear optical
properties of the EIT system. In Sec. V, we discuss the response
of the EIT medium to nonclassical light fields. Finally, we
summarize our results and make concluding remarks in Sec. VI.

%%%%%%%%%%%%%%%%%%%%%%%%%%%%%%%%%%%%%%%%%%%%%%
\section {The model and adiabatic solution}
%%%%%%%%%%%%%%%%%%%%%%%%%%%%%%%%%%%%%%%%%%%%%%
 Let us consider a three-level
atom, with energy levels $E_1<E_3<E_2$,
interacting with two quantized laser fields,
in the $\Lambda$-type configuration (see Fig. 1):
The lower two levels $|1\rangle$ and $|3\rangle$
are coupled to the upper level $|2\rangle$.
Going over to an interaction picture with respect to
$H_0=\sum_m E_m |m\rangle \langle m|- \hbar[\Delta_1
(|2\rangle \langle 2| + |3\rangle \langle 3|)
+\omega_1\hat{a}^\dagger_1 \hat{a}_1
+\omega_2\hat{a}^\dagger_2 \hat{a}_2]$,
under the rotating-wave approximation, we obtain  the total Hamiltonian of the system  in the form
\begin{eqnarray}
\hat{H}&=&\hbar\Delta_1|2\rangle\langle 2| + \hbar(\Delta_1-\Delta_2)|3\rangle\langle3|) \nonumber \\
\label{e1}
& &- \hbar(g_1\hat{a}_1|2\rangle\langle 1|+g_2\hat{a}_2|2\rangle\langle 3| + H.c.  ),
\end{eqnarray}
where $|m\rangle$ ($m=1,2,3$) are atomic
sates, $\hat{a}_j$ and $\hat{a}^\dagger_j$
($j=1,2$) the annihilation and creation
operators of the probe and coupling laser
modes, and the two  coupling constants are defined by
$g_1=\mu_{21}{\cal E}_1/\hbar$ and $g_2=\mu_{23}{\cal E}_2/\hbar$ with
$\mu_{ij}$ denoting   a  transition dipole-matrix element
between states $|i\rangle$ and $|j\rangle$,
${\cal E}_i=\sqrt{\hbar\omega_i/2\epsilon_0V}$
being the electric field per photon for light of frequency $\omega_i$
in a mode volume $V$.

Eigenvalues of above Hamiltonian can be  generally expressed as
the following form
\begin{eqnarray}
\label{e2}
E^{(+)}_{n_1, n_2}&=&\left [-\frac{q}{2}
+\sqrt{\left (\frac{q}{2}\right )^2+\left (\frac{p}{3}\right )^3}\right ]^{1/3}\nonumber \\
& &+\left [-\frac{q}{2}-\sqrt{\left (\frac{q}{2}\right )^2+\left (\frac{p}{3}\right )^3}\right ]^{1/3}-\frac{1}{3}A, \nonumber \\
E^{(-)}_{n_1, n_2}&=&\Lambda_1\left [-\frac{q}{2}
+\sqrt{\left (\frac{q}{2}\right )^2+\left (\frac{p}{3}\right )^3}\right ]^{1/3}\nonumber \\
& &+\Lambda_2\left [-\frac{q}{2}-\sqrt{\left (\frac{q}{2}\right )^2+\left (\frac{p}{3}\right )^3}\right ]^{1/3}-\frac{1}{3}A,\\
E^{(0)}_{n_1, n_2}&=&\Lambda_2\left [-\frac{q}{2}
+\sqrt{\left (\frac{q}{2}\right )^2+\left (\frac{p}{3}\right )^3}\right ]^{1/3}\nonumber \\
& &+\Lambda_1\left [-\frac{q}{2}-\sqrt{\left (\frac{q}{2}\right )^2+\left (\frac{p}{3}\right )^3}\right ]^{1/3}-\frac{1}{3}A, \nonumber
\end{eqnarray}
where the two constants $\Lambda_1=\exp(i2\pi/3)$ and $\Lambda_2=\exp(-i2\pi/3)$, other parameters are given by
\begin{equation}
\label{e3}
p=B-\frac{1}{3}A^2, \hspace{0.5cm}q=C-\frac{1}{3}AB+\frac{2}{27}A^3,
\end{equation}
with
\begin{eqnarray}
A&=&-2\Delta_1+\Delta_2, \hspace{0.5cm}
C=g^2_1n_1(\Delta_1-\Delta_2), \nonumber \\
B&=&\Delta_1(\Delta_1-\Delta_2)-g^2_2(n_2+1)-g^2_1n_1.
\end{eqnarray}

%%%%%%%%%%%%%%%%%%%%%%%%%%%%%%%%%%%%%%%%%%%%%%%%%%%%%%%%%%%%%
\begin{figure}[htb]
\begin{center}
\includegraphics[width=7cm]{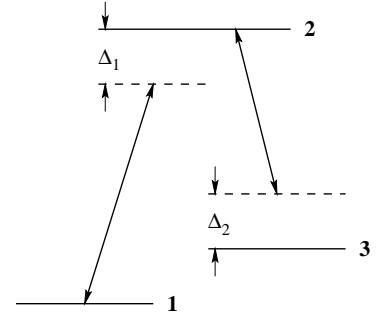}
\end{center}
\vskip 0.2cm \caption{A sketch of the atomic energy levels coupled
to
 the coupling and probe fields.}
\label{fig1}
\end{figure}
%%%%%%%%%%%%%%%%%%%%%%%%%%%%%%%%%%%%%%%%%%%%%%%%%%%%%%%%%%%%%

The three eigenvalues $E^{(+)}_{n_1, n_2}$,  $E^{(-)}_{n_1, n_2}$
and  $E^{(0)}_{n_1, n_2}$ are unequal and real when $p$ and $q$ obey
the inequality  $q^2/4+p^3/27 <0$.

Since we are concerned with behaviors of the system in the vicinity of
 resonance, assuming the two detunings $\Delta_1$ and  $\Delta_2$ are small,
we can develop a perturbation theory in which we retain only terms
linear in
 $\Delta_1$ and $\Delta_2$. Then eigenvalues in Eq. (\ref{e2}) become
the following simple expressions
\begin{eqnarray}
E^{(\pm)}_{n_1, n_2}&=&\frac{\Omega^2_1
+ 2\Omega^2_2}{2\Omega^2}\Delta_1 -\frac{\Omega^2_2}{2\Omega^2}\Delta_2
\pm\frac{\Omega}{2}, \nonumber \\
\label{e5}
E^{(0)}_{n_1, n_2}&=&\frac{\Omega^2_1}
{\Omega^2}(\Delta_1-\Delta_2),
\end{eqnarray}
where  we have defined  $\Omega_1=2g_1\sqrt{n_1}$,
$\Omega_2=2g_2\sqrt{n_2+1}$ and
$\Omega=\sqrt{\Omega^2_1+\Omega^2_2}$.

Eigenstates of the Hamiltonian (\ref{e1}) are  called  dressed
states. Dressed states corresponding to above eigenvalues
 have the following form:
\begin{eqnarray}
\label{e6}
|\phi^{(i)}_{n_1, n_2}\rangle&=&
a_i |1,n_1, n_2\rangle
+b_i |2,n_1-1, n_2\rangle
\nonumber \\
& &  +c_i |3,n_1-1, n_2+1\rangle,
\end{eqnarray}
where $i=\pm, 0$. Up to first-order perturbation theory in the detunings
$\Delta_1$ and $\Delta_1$, the coefficients in Eq. (\ref{e6})
are given by
\begin{eqnarray}
\label{e7}
a_{\pm} &=&-\frac{\Omega_1}
{\sqrt2\Omega} \left [1\mp
\frac{\Omega^2_1 + 4\Omega^2_2}
{2\Omega^3}\Delta_1 \pm \frac{3\Omega^2_2}
{2\Omega^3}\Delta_2  \right ],
\nonumber \\
b_{\pm} &=&\pm\frac{1}{\sqrt2}
\left [1\pm \frac{\Omega^2_1}{2\Omega^3}
\Delta_1 \pm \frac{\Omega^2_2}
{2\Omega^3}\right ], \\
c_{\pm} &=&-\frac{\Omega_2}
{\sqrt2\Omega} \left [1\pm
\frac{3\Omega^2_1}{2\Omega^3}\Delta_1 \mp
\frac{4\Omega^2_1+3\Omega^2_2}
{2\Omega^3}\Delta_2\right ],
\nonumber
\end{eqnarray}
and
\begin{eqnarray}
\label{e8}
a_{0}=-\frac{\Omega_2}{\Omega},
\;\;
b_{0} =\frac{2\Omega_1\Omega_2}
{\Omega^3}(\Delta_1-\Delta_2),\;\;
c_{0} =\frac{\Omega_1}{\Omega}.
\end{eqnarray}

The complete set of dressed states for
the system under consideration comprises
the states $|\phi^{(i)}_{n_1,n_2}\rangle
(i=\pm, 0)$ for each $n_1>0$, and  $n_2 \ge 0$ together
with other two states $|1,0, n_2\rangle$
and $|3,n_1,0\rangle$ with zero eigenvalues.
Knowing this set allows us to determine the
time evolution of the atom-field system for
any initial configuration.

We assume that the atom is initially in the
ground state, while the coupling and probe
lasers in a coherent state $|\alpha,
\beta\rangle$, with $\alpha$ and $\beta$
being supposed to be real for simplicity. Namely
the initial state of the atom-field system
is assumed to be
\begin{equation}
\label{e9}
|\Psi(0)\rangle =|1\rangle\otimes |\alpha,\beta\rangle
\equiv \sum^{\infty}_{n_1,n_2=0}C_{n_1,n_2}|1,n_1,n_2\rangle,
\end{equation}
where the coefficient $C_{n_1,n_2}$ is given by
\begin{equation}
\label{e9a}
C_{n_1,n_2}=\exp[-(\alpha^2+\beta^2)/2]
\frac{\alpha^{n_1}\beta^{n_2}}{\sqrt{n_1!n_2!}}.
\end{equation}

Eq. (\ref{e9}) indicates that there is no atom-photon and
photon-photon entanglement
in the initial state of the atom-field system.
In the usual treatment with both lasers
being classical external fields, one is
concerned with a steady state of the
atomic system. The counterpart of the
steady state in our fully quantum treatment,
according to the commonly used {\it adiabatic
hypothesis} in quantum scattering theory
and quantum field theory\cite{gell}, is
the state that evolves from the initial
state (\ref{e9}) with the couplings $g_1$ and
$g_2$ {\it adiabatically turned on}.
Physically this is equivalent to having
localized laser pulses before they enter
the atomic vapor, with the pulse shape
sufficiently smooth. In conformity to the
Adiabatic Theorem, we need to identify
a linear combination of the dressed
states (\ref{e2}) that tends to the initial
state (\ref{e9}) if we take the limits $g_1,g_2
\to 0$ (or $\Omega_1, \Omega_2 \to 0$).
According to Eqs. (\ref{e3}-\ref{e5}), the ordering of
the limits $\Omega_1 \to 0$ and $\Omega_2
\to 0$ is important. Corresponding to
the actual conditions in which EIT is
observed, the correct ordering is first
$\Omega_1 \to 0$ and then $\Omega_2 \to 0$.
In our interaction picture, this procedure
selects the $i=0$ state in Eq. (\ref{e6}).
Transforming it back to the Schr\"{o}dinger
picture, we identify the following state
as the state that evolves adiabatically
from the initial state:
\begin{equation}
\label{e10}
|\Psi(t)\rangle=\sum^{\infty}_{n_1,n_2=0}
C_{n_1,n_2} \exp\left [-ie^{(0)}_{n_1,n_2}t\right ]
|\phi^{(0)}_{n_1,n_2}\rangle,
\end{equation}
where we have introduced the following notation
\begin{equation}
\label{e10a}
e^{(0)}_{n_1,n_2}=E_1+\omega_1n_1+\omega_2n_2+E^{(0)}_{n_1,n_2}.
\end{equation}

The  state given in Eq. (\ref{e10}) can be viewed as the counterpart
of the usual "steady" state in our treatment.
We see that the population of the upper level
$|2\rangle$ is zero up to first order in detunings
$\Delta_1$ and $\Delta_2$. This means that there is no
absorption, implying the phenomenon of EIT. This is the adiabatic EIT.
In the next section we will see  that for transient-state  EIT
to occur the usual treatment
requires the coupling laser be much stronger
than the probe laser, i.e.,  $\Omega_2
\gg \Omega_1$; then most atoms are populated in
the ground state. The adiabatic EIT does not require
this, so broader conditions are allowed:
The coupling and probe lasers can be equally
strong or even the probe laser is stronger.
Then atoms can be in a more general superposition
of two atomic states, which are entangled with
the coupling and probe fields.

From Eq. (\ref{e10}) one can get  the total density
operator of the atom-field system. After
taking trace over the field states, one
finds  the atomic reduced density operator to be
\begin{eqnarray}
\label{e11}
\rho^A(t)&=&\sum^{\infty}_{n_1,n_2=0}\left \{|D_1(n_1,n_2)|^2
|1\rangle\langle 1| \right.  \nonumber \\
& &+|D_2(n_1+1,n_2)|^2|2\rangle\langle 2| \nonumber \\
& &+|D_3(n_1+1,n_2-1)|^2|3\rangle\langle 3|\nonumber \\
& &+\left [D_1(n_1,n_2)D^*_2(n_1+1,n_2)|1\rangle\langle 2| \right.
  \nonumber \\
& &+D_1(n_1,n_2)D^*_3(n_1+1,n_2-1)|1\rangle\langle 3|\nonumber \\
& &+ D_2(n_1+1,n_2)D^*_3(n_1+1,n_2-1)|2\rangle\langle 3|\nonumber \\
& &\left. \left.+ H.c.\right ]\right \},
\end{eqnarray}
where  D-coefficients are defined by
\begin{eqnarray}
\label{e12}
D_1(n_1,n_2)&=&a_0C_{n_1,n_2}\exp\left [-ie^{(0)}_{n_1,n_2}t\right ],
 \nonumber \\
D_2(n_1,n_2)&=&b_0C_{n_1,n_2}\exp\left [-ie^{(0)}_{n_1,n_2}t\right ], \\
D_3(n_1,n_2)&=&c_0C_{n_1,n_2}\exp\left [-ie^{(0)}_{n_1,n_2}t\right ], \nonumber
\end{eqnarray}
where $a_0$, $b_0$, abd $c_0$ are given by Eq. (\ref{e8}).

%%%%%%%%%%%%%%%%%%%%%%%%%%%%%%%%%%%%%%%%%%%%%%%%%%%%

\section{Dispersion and group velocity}

%%%%%%%%%%%%%%%%%%%%%%%%%%%%%%%%%%%%%%%%%%%%%%%%%%%%
When an electromagnetic wave of frequency $\nu$ propagates through a
nonabsorptive but dispersive medium, the phase velocity is $c/n$ with
$c$ being the speed of light in vacuum, and  $n$  the refractive index
which is related to susceptibility of the medium $\chi$  by the
relation $n=\sqrt{1+\chi}$,
while the group velocity is $c/[n+\nu(dn/d\nu)]$.
So long as $dn/d\nu>0$ which corresponds to normal dispersive,
the group velocity is less than the phase velocity. In what follows
we shall indicate that under EIT conditions both probe and coupling
lasers exhibit normal dispersion, and their group velocities are greatly
reduced while they have the same  phase velocity $c$.

In order to get the susceptibility of the medium, we need to  calculate
the polarization of the atomic medium which is related to off-diagonal elements
of the atomic  reduced density operator through the following relation:
\begin{equation}
\label{e13}
P=N(\mu_{12}\rho^A_{21}+\mu_{32}\rho^A_{23})+ c.c.,
\end{equation}
where  $N$ is the number density of atoms,
$\mu_{ij}$ denotes  a  transition dipole-matrix element
between states $|i\rangle$ and $|j\rangle$.

Transferring the induced polarization of the atomic medium
 $P$ to a Fourier  representation of a frequency $\omega$,
one can define the susceptibility of the medium $\chi(\omega)$ by
\begin{equation}
\label{e14}
P(\omega)=\epsilon_0\chi_s(\omega)E(\omega),
\end{equation}
where $E(\omega)$ is the Fourier component of the mean value of
the total electric field at the frequency $\omega$, $\epsilon_0$
the free space permittivity.

From Eqs. (\ref{e11}) and (\ref{e12}) we get the optical coherence of
the system at time $t$
\begin{eqnarray}
\label{e15a}
\rho^A_{21}(t)&=&\sum^{\infty}_{n_1,n_2=0}P(n_1,n_2)
                 a^*_0(n_1,n_2)b_0(n_1+1,n_2) \nonumber \\
& &\times\frac{\alpha}{\sqrt{n_1+1}}\exp\left \{i\left [e^{(0)}_{n_1,n_2}
-e^{(0)}_{n_1+1,n_2}\right ]t\right\},  \\
\rho^A_{23}(t)&=&\sum^{\infty}_{n_1,n_2=0}P(n_1,n_2)
               b_0(n_1+1,n_2)c^*_0(n_1+1,n_2-1)  \nonumber \\
& &\times\frac{\sqrt{n_2}|\alpha|^2}{(n_1+1)\beta^*}  \exp\left \{i\left [e^{(0)}_{n_1+1,n_2-1}
-e^{(0)}_{n_1+1+1,n_2}\right ]t\right\}, \nonumber
\end{eqnarray}
where the weighting factor is given by
\begin{equation}
\label{e15b}
P(n_1,n_2)=\frac{\bar{n}^{n_1}_{\alpha}\bar{n}^{n_2}_{\beta}}
{n_1!n_2!}e^{-(\bar{n}_{\alpha}+\bar{n}_{\beta})},
\end{equation}
with $\bar{n}_{\alpha}=|\alpha|^2$ and $\bar{n}_{\beta}=|\beta|^2$,
respectively,  being the initial mean photon numbers of
the probe and coupling
lasers in the coherent state $|\alpha, \beta\rangle$.

In the case $\bar{n}_{\alpha} \gg 1$ and $\bar{n}_{\beta} \gg 1$
($\bar{n}\sim 10^4$  for  a recent
experiment of light storage \cite{liu}, $\bar{n}\sim 10^4$),
 called the large-$n$ limit below,  the summation over $n_1$  and $n_2$ in
Eq. (\ref{e15a}) can be performed approximately. If one notices that
 the weighting function  $P(n_1,n_2)$ will peak at values
 $n_1=\bar{n}_{\alpha}$ and $n_2=\bar{n}_{\beta}$ with
relatively narrow dispersion:
$(\langle n^2_{\alpha}\rangle -\bar{n}^2_{\alpha})^{1/2}
=\sqrt{\bar{n}_{\alpha}}$
and $(\langle n^2_{\beta}\rangle -\bar{n}^2_{\beta})^{1/2}
=\sqrt{\bar{n}_{\beta}}$,
this implies  that the photon
distributions are sharply peaked around  their mean values.
So the rapid oscillations in the time record in Eq. (\ref{e15a}) are
just dominant oscillations which occur for $n_1=\bar{n}_{\alpha}$
and  $n_2=\bar{n}_{\beta}$, i.e.,
$\Omega_1=\bar{\Omega}_1(n_1=\bar{n}_{\alpha})$
and $\Omega_2=\bar{\Omega}_2(n_2=\bar{n}_{\beta})$.
Indeed we can expand  the Rabi frequencies $\Omega_1=2g_1\sqrt{n_1}$ and
 $\Omega_2=2g_2\sqrt{n_2+1}$ about $\bar{n}_{\alpha}$
and $\bar{n}_{\beta}$ to obtain
\begin{eqnarray}
\label{e15c}
2g_1\sqrt{n_1}&=&2g_1\sqrt{\bar{n}_{\alpha}}
+\frac{1}{\sqrt{\bar{n}_{\alpha}}}(n_1-\bar{n}_{\alpha})
\nonumber \\
& &-\frac{1}{4\bar{n}^{3/2}_{\alpha}}(n_1-\bar{n}_{\alpha})^2 + \cdots
\nonumber  \\
 2g_2\sqrt{n_2+1}&=&2g_2\sqrt{\bar{n}_{\beta}+1}
+\frac{1}{\sqrt{\bar{n}_{\beta}}}(n_2-\bar{n}_{\beta})
\nonumber \\
& &-\frac{1}{4\bar{n}^{3/2}_{\beta}}(n_1-\bar{n}_{\beta})^2 + \cdots
\end{eqnarray}
To the lowest order of $n_1-\bar{n}_{\alpha}$ and $n_2-\bar{n}_{\beta}$,
we get $\Omega_1=\bar{\Omega}_1(\bar{n}_{\alpha})$
and $\Omega_2=\bar{\Omega}_2(\bar{n}_{\beta})$.

After substituting the lowest-order values of the Rabi frequencies,
$a_0(n_1,n_2)$, $b_0(n_1,n_2)$, and $e^{(0)}_{n_1,n_2}$ in Eq.
(\ref{e15a})  are  independent of $n_1$ and $n_2$.
The summations  over $n_1$ and $n_2$   in Eq.
(\ref{e15a})  are  carried out only for $P(n_1,n_2)/\sqrt{n_1+1}$ and
$P(n_1,n_2)\sqrt{n_2}/(n_1+1)$. These can be done easily. Then
we obtain  the Fourier component  of the optical coherence of the
adiabatically prepared state at the frequency
$\omega$
\begin{eqnarray}
\label{e15}
\rho^A_{21}(\omega)&=&\bar{a}_0\bar{b}_0[\delta(\omega-\omega_1)+\delta(\omega+\omega_1)], \nonumber \\
\rho^A_{23}(\omega)&=&\bar{c}_0\bar{b}_0 [\delta(\omega-\omega_2)+\delta(\omega+\omega_2)],
\end{eqnarray}
where $\bar{a}_0=a_0(\bar{n}_{\alpha},\bar{n}_{\beta})$,
$\bar{b}_0=b_0(\bar{n}_{\alpha},\bar{n}_{\beta})$,
and $\bar{c}_0=c_0(\bar{n}_{\alpha},\bar{n}_{\beta})$.

The Fourier component of the mean value  of the total electric
field $E(\omega)$ can be calculated from the adiabatically prepared  state
(\ref{e10}).  In the large $n$ approximation, we find that
\begin{eqnarray}
\label{e16}
E(\omega)&=&{\cal E}_1\alpha[\delta(\omega-\omega_1)
+\delta(\omega+\omega_1)]\nonumber \\
& &+{\cal E}_2\beta[\delta(\omega-\omega_2)
+\delta(\omega+\omega_2)],
\end{eqnarray}
where ${\cal E}_i
=(\hbar\omega_i/2\epsilon_0V)^{1/2} (i=1,2) $ with  $V$ the
quantized volume.

From Eq. (\ref{e16}) we obtain $E(\omega_1)={\cal E}_1\alpha$ and
$E(\omega_2)={\cal E}_2\beta$, which imply that the mean values of
the electric fields are unchanged when the pulses enter the
medium. This is in agreement with the argument in Ref. \cite{hh}
for the case of the classical probe and coupling laser fields.
Since the power density is also unchnged  as the pulses enter the
medium, the energy density in the medium must increase. It is this
increase that leads the spatial compression of a light pulse in
the medium.

We first consider the propagation of the probe laser. For the
frequency of the probe laser $\omega_1$, substituting Eqs.
(\ref{e15}) and (\ref{e16}) into Eqs. (\ref{e13}) and (\ref{e14})
and retaining only terms linear in the detunings  $\Delta_1$ and
$\Delta_2$ we obtain the following expression of the
susceptibility
\begin{equation}
\label{e17}
\chi_s(\omega_1)=-\frac{4N|\mu_{12}|^2
\bar{\Omega}^2_2(\Delta_1-\Delta_2)}{\hbar\epsilon_0
(\bar{\Omega}^2_1+\bar{\Omega}^2_2)^2},
\end{equation}
where $\bar{\Omega}_1=\Omega_1(\bar{n}_{\alpha}, \bar{n}_{\beta})$
and $\bar{\Omega}_2=\Omega_2 (\bar{n}_{\alpha},\bar{n}_{\beta})$
are the Rabi frequencies of the coupling and probe lasers,
respectively. It should be pointed out that the expression
(\ref{e17}) is just a perturbation result of the susceptibility
for the small detunigs up to the first order of $\Delta_1$ and
$\Delta_2$. Eq. (\ref{e17}) indicates that the imaginary part of
the susceptibility vanishes under the first order approximation of
the detunings.  When the EIT occurs,we have $\Delta_1=0=\Delta_2$.
Then both real and imaginary parts of the susceptibility are zero.
Hence, the Kramers-Kronig relation is not violated. When the
coupling laser is on resonance, Eq. (\ref{e17}) reduces to the
result in our previous paper \cite{kua}. It is interesting to note
that this result is valid for arbitrary ratio of  $\bar{\Omega}_1/
\bar{\Omega}_2$.

We can see that Eq. (\ref{e17}) exhibits the signature of EIT:
the linear susceptibility vanishes
at the resonance ($\Delta_1=0=\Delta_2$), which implies that
the probe laser has the phase velocity $c$.
From Eq. (\ref{e17}) we can also know  that
at resonance,   $d\chi_s/d\omega_1>0$. Hence we can conclude
that  the medium exhibits  the normal dispersion.

A dispersive medium can be characterized by a light  group velocity.
Steep dispersion is associated with larger modifications of the group
velocity. At the resonance frequency the group velocity of
the probe laser pulse is related to the derivative of  the susceptibility
through the relation:$v^{(s)}_{pg}=c/[1+(\omega_1/2)(d\chi_s/d\omega_1)]$,
where the derivative of $\chi_s (\omega_1)$  is evaluated at
 the probe frequency $\omega_1$ with $c$ the  speed of light in vacuum.
Thus we have
\begin{equation}
\label{e18}
v^{(s)}_{pg}=v^0_{pg}\frac{\left (\bar{\Omega}^2_1
+\bar{\Omega}^2_2 \right )^2 }
{\bar{\Omega}^4_2},
\end{equation}
where $v^0_{pg}$  is the usual expression for the group
velocity \cite{hau} given by
\begin{equation}
\label{e19}
v^0_{pg}=\frac{\hbar c\epsilon_0
\bar{\Omega}^2_2}{2\omega_1|\mu_{12}|^2N}.
\end{equation}

Eq. (\ref{e18}) indicates  that in general the group velocity
of the probe laser depends on the Rabi
frequency of the coupling as well
as the probe laser. The $\bar{\Omega}_1$
dependence of the group velocity is a new
result of the fully quantum  formalism, closely related
to the higher order nonlinear
susceptibilities we will discuss below.

These general results should be compared with earlier
studies of EIT-based dispersive properties
\cite{harri,flescu}. In order to do this, we  expand the susceptibility
given in Eq. (\ref{e17}) in terms of the powers
of $ \bar{\Omega}_1/ \bar{\Omega}_2$. When
$\bar{\Omega}_1 \ll \bar{\Omega}_2$, the
first term of this expansion gives us the
linear susceptibility
\begin{equation}
\label{e20}
\chi^{(1)}_s(\omega_1)=-\frac{4N|\mu_{12}|^2
(\Delta_1-\Delta_2)}{\hbar\epsilon_0\bar{\Omega}^2_2},
\end{equation}
which is the same as Eq. (16a) in Ref. \cite{flescu}.
Note that the Rabi frequencies here are twice of those
in Ref. \cite{flescu} by definition.

On the other hand,  when the probe and coupling lasers are on resonance,
i.e., $\Delta_1=0=\Delta_2$,
from Eq. (\ref{e20}) we find that
\begin{eqnarray}
\label{e20a}
\chi^{(1)}_s(\omega_1)&=&0, \hspace{0.3cm}
\frac{d\chi^{(1)}_s}{d\omega_1}=\frac{4|\mu_{12}|^2N}
{\hbar\epsilon_0\bar{\Omega}^2_2},
\nonumber \\
 \frac{d^2\chi^{(1)}_s}{d\omega^2_1}&=&0.
\end{eqnarray}
Thus, at the lowest order we recover results
in Ref. \cite{harri} (see Table 1 in  \cite{harri}), when the decay rates
of states are neglected at the resonant
frequency of the probe laser. Similarly,
keeping only the lowest-order  term  in
$\bar{\Omega}_1/\bar{\Omega}_2$, Eq. (\ref{e18})
will essentially  give the same  expression for the group
velocity as reported in Ref. \cite{hau}, i.e., the expression
 given by Eq. (\ref{e19}), which does not depend
on $\bar{\Omega}_1$ and involves the
contributions only from linear susceptibility.

The refractive index of the medium  can be generally obtained
 from the susceptibility  (\ref{e17}) by definition
$n=\sqrt{1+ \chi}$.
 Making use of Eqs. (\ref{e17}) and (\ref{e18}),
we find that  the  refractive index change in the vicinity of
the resonance  can be expressed in terms of the group velocity of the probe
laser as
\begin{equation}
\label{e21} \Delta n_s(\omega_1)=\frac{\lambda_1}{2\pi}
\frac{(\Delta_1-\Delta_2)}{v^{(s)}_{pg}}.
\end{equation}

It is worthwhile to note that this  refractive-index change
involves the contributions of all orders of nonlinear
susceptibilities due to the intensity dependence of the group
velocity in (\ref{e18}).  For the slow light experiment of
$v^{(s)}_{pg}=17$ m/s with parameters \cite{hau}
$\Delta_1=1.3\times 10^6$ rad/s, $\Delta_2=0$,  and
$\lambda_1=589$ nm, from Eq.(\ref{e21}) we obtain $\Delta
n_s(\omega_1)=8.2\times 10^{-3}$. This value is the same as that
indirectly measured  in Ref. \cite{hau}. It should be pointed out
that the much larger change of refractive-index coefficients or
the much larger normal dispersive of the medium which leads to
much slower group velocity, is the direct result of EIT but not
the result of strong coupling laser. In fact, a strong coupling
laser is unnecessary for the adiabatic EIT situation.

We then consider the propagation of the coupling laser. For the frequency
of the coupling laser $\omega_2$,  from Eqs. (\ref{e14}), and (\ref{e15})
 we obtain the following susceptibility
\begin{equation}
\label{e22}
\chi_s(\omega_2)=\frac{4N|\mu_{32}|^2
\bar{\Omega}^2_1(\Delta_1-\Delta_2)}{\hbar\epsilon_0
(\bar{\Omega}^2_1+\bar{\Omega}^2_2)^2}.
\end{equation}

Eq. (\ref{e22}) indicates that at resonance, we have $\chi_s(\omega_2)=0$,
and $d\chi_s(\omega_2)/d\omega_2>0$. This means that the coupling laser has
the phase velocity $c$ and exhibits the normal dispersion.

 When $\bar{\Omega}_1 \ll \bar{\Omega}_2$,
 we expand the susceptibility
given in Eq. (\ref{e22}) in terms of powers
of $ \bar{\Omega}_1/ \bar{\Omega}_2$.
The  first term of this expansion gives us the
linear susceptibility of the coupling laser
\begin{equation}
\label{e23}
\chi^{(1)}_s(\omega_2)=\frac{4N|\mu_{32}|^2
\bar{\Omega}^2_1(\Delta_1-\Delta_2)}{\hbar\epsilon_0
\bar{\Omega}^4_2}.
\end{equation}
Thus, at the lowest order we recover results
in Ref. \cite{flescu} (Eq.(B2) in Ref. \cite{flescu}).

Making use of the susceptibility (\ref{e22}), we obtain
the group velocity for the coupling  laser pulse
\begin{equation}
\label{e24}
v^{(s)}_{cg}=v^0_{cg}\frac{(\bar{\Omega}^2_1+\bar{\Omega}^2_2)^2}
{\bar{\Omega}^4_1},
\end{equation}
where $v^0_{cg}$ is given by
\begin{equation}
\label{e25}
v^0_{cg}=\frac{\hbar c\epsilon_0
\bar{\Omega}^2_1}{2\omega_2|\mu_{32}|^2N}
=v^0_{pg}\frac{\omega_1I_1|\mu_{12}|^2}{\omega_2I_2|\mu_{32}|^2}.
\end{equation}
where $ v^0_{pg}$ has given in Eq. (\ref{e19}),
 $I_1=2\epsilon_0c\alpha^2$ and   $I_2=2\epsilon_0c\beta^2$ are
the intensities of the incident probe and  coupling lasers, respectively.
 For the slow light experiment of
$v^{(s)}_{pg}=17$ m/s with parameters \cite{hau}
$I_1=1$ mW/cm$^2$, $I_2=40$ mW/cm$^2$,  $\omega_1/\omega_2\approx 1$ and
$\mu_{12}/\mu_{32}\approx 1.22$,
from Eqs. (\ref{e24}) and (\ref{e25}) we obtain
$v^{(s)}_{cg}\approx 1020$ m/s. Hence the group velocity of the coupling
laser is greatly reduced, although it is much greater than that of the
probe laser.

The  refractive-index change in the vicinity of
the resonance is given by
\begin{equation}
\label{e26a}
\Delta n_s(\omega_2)=\frac{2N|\mu_{32}|^2
\bar{\Omega}^2_1(\Delta_1-\Delta_2)}{\hbar\epsilon_0
(\bar{\Omega}^2_1+\bar{\Omega}^2_2)^2},
\end{equation}
which can be expressed as the following simple form
\begin{equation}
\label{e26}
\Delta n_s(\omega_2)=\frac{\lambda_2}{2\pi}
\frac{(\Delta_1-\Delta_2)}{v^{(s)}_{cg}}.
\end{equation}
 For the slow light experiment of
$v^{(s)}_{pg}=17$ m/s with parameters \cite{hau}
$\Delta_1=1.3\times 10^6$ rad/s, $\Delta_2=0$,  and $\lambda_2\approx 589$ nm,
from Eq.(\ref{e26}) we obtain $\Delta n_s(\omega_2)=2.1\times 10^{-4}$.
This value is less one order of magnitude  than that of the probe laser
$\Delta n_s(\omega_1)=8.2\times 10^{-3}$.

%Nonlinearities play an important role not only
%in nonlinear optics but also in quantum optics.
%They may be used for generation of enhanced
%squeezing \cite{tom}, quantum computation and
%quantum teleportation\cite{vit}, and quantum
%nondemolition measurements \cite{bra}.
%Since EIT takes place in the vicinity of
%atomic resonance, large nonlinearities are
%naturally expected.

%%%%%%%%%%%%%%%%%%%%%%%%%%%%%%%%%%%%%%%%%%%%%%%

\section{Nonlinear  refractive indices}

%%%%%%%%%%%%%%%%%%%%%%%%%%%%%%%%%%%%%%%%%%%%%%%

We note that in the conventional approach,
it is difficult to obtain nonlinear susceptibilities
higher than $\chi^{(1)}$ and $\chi^{(3)}$.
However, in our formalism we can get arbitrary
higher-order nonlinear susceptibilities once
for all. For the propagation of the probe laser it is easy to see  that
the higher-order terms of the expansion in Eq. (\ref{e17})
give rise to higher-order nonlinear susceptibilities defined by
\begin{eqnarray}
\label{e27}
\chi_s(\omega_1)&=&\chi^{(1)}_s(\omega_1)
+\chi^{(3)}_s(\omega_1)|E(\omega_1)|^2 \nonumber \\
& &+\chi^{(5)}_s(\omega_1)|E(\omega_1)|^4
+ \chi^{(7)}_s(\omega_1)|E(\omega_1)|^6 +\cdots
\end{eqnarray}
where $\chi^{(1)}_s(\omega_1)$ is the linear
susceptibility given in Eq. (\ref{e20}), and
$\chi^{(k)}_s(\omega_1) (k \ge 3)$ represent
the $k$th-order nonlinear susceptibility.

Again, assuming that the coupling laser is
stronger than the probe, we expand the
susceptibility (\ref{e17}) in powers of
$\bar{\Omega}_1/\bar{\Omega}_2$. Rewriting
this expansion in the form of Eq. (\ref{e27}), we
find
\begin{eqnarray}
\label{e28}
\chi^{(3)}_s(\omega_1)&=&\frac{8\hbar\epsilon_0c^2|\mu_{12}|^4N}
{|\mu_{32}|^4I^2_2}(\Delta_1-\Delta_2),
\nonumber \\
\chi^{(5)}_s(\omega_1)&=&-\frac{24\hbar\epsilon^2_0c^3|\mu_{12}|^6N}
{|\mu_{32}|^6I^3_2}(\Delta_1-\Delta_2), \\
\chi^{(7)}_s(\omega_1)&=&\frac{192\hbar\epsilon^3_0c^4|\mu_{12}|^8N}
{3|\mu_{32}|^8I^4_2}(\Delta_1-\Delta_2), \nonumber
\end{eqnarray}
where $\chi^{(1)}_s(\omega_1)$ is given by Eq.
(\ref{e20}).

A light wave propagating through a nonlinear medium collects
a nonlinear phase shift which is described by the expansion
of the refractive index on powers of the light intensity  \cite{rat}:
\begin{eqnarray}
\label{e29}
n_s&=&n^{(s)}_0+n^{(s)}_2|E_1|^2+n^{(s)}_4|E_1|^4 +n^{(s)}_6|E_1|^6 + \cdots
\end{eqnarray}
where the first nonlinear correction to the
refractive index is the Kerr coefficient
$n^{(s)}_2$, which is related to
$\chi^{(3)}_s(\omega_1)$.
$n^{(s)}_k (k\ge 4)$ are higher-order nonlinear
refractive-index coefficients, related to
higher-order nonlinear susceptibilities
up to $\chi^{(k+1)}_s(\omega_1)$.

Making use of Eq. (\ref{e27}), we find that the linear and nonlinear
refractive-index coefficients can be expressed in terms of the linear
and nonlinear susceptibilities such as
\begin{eqnarray}
\label{e30}
n^{(s)}_0(\omega)&=&1+\chi^{(1)}_s(\omega), \hspace{0.3cm}
n^{(s)}_2(\omega)=\frac{1}{2n_0}\chi^{(3)}_s(\omega), \nonumber \\
n^{(s)}_4(\omega)&=&\frac{1}{2n_0}\chi^{(5)}_s(\omega)-
                   \frac{1}{8n^3_0}\left (\chi^{(3)}_s(\omega)\right )^2, \\
n^{(s)}_6(\omega)&=&\frac{1}{2n_0}\chi^{(7)}_s(\omega)-
                   \frac{1}{4n^3_0}\chi^{(3)}_s(\omega)\chi^{(5)}_s(\omega)
                   \nonumber \\
& & +\frac{1}{16n^5_0}\left (\chi^{(3)}_s(\omega)\right )^3. \nonumber
\end{eqnarray}

Making use of Eq. (\ref{e28}), we obtain
from Eq. (\ref{e30}) nonlinear refractive index
coefficients:
\begin{eqnarray}
\label{e31}
n^{(s)}_2&=&-\frac{2\epsilon_0c}{I_2}\chi^{(1)}_s(\omega_1)
=\frac{2\epsilon_0c(\Delta_1-\Delta_2)}{\pi I_2} \frac{\lambda_1}{v^0_{pg}},
\nonumber \\
n^{(s)}_4&=&-\frac{3\epsilon_0cn^{(s)}_2}{I_2}, \hspace{0.5cm}
n^{(s)}_6=-\frac{8\epsilon_0cn^{(s)}_4}{3I_2}.
\end{eqnarray}

Note that  the sign of $n^{(s)}_i$ is of significance. In
particular, the sign of $n^{(s)}_2$ can be used to  examine  the
self-focusing/-defocusing effect of the medium. When
$n^{(s)}_2>0$, the medium is  self-focusing.  When  $n^{(s)}_2<0$,
the medium is  self-defocusing. Eq. (\ref{e31}) tell us that one
can manipulate  the self-focusing/-defocusing effect of the medium
by changing the relative detuning between the coupling and probe
lasers.

From Eqs. (\ref{e28}) and (\ref{e31}) we can see that
to increase the value of nonlinear refractive-index
coefficients one may either increase the atomic
density or decrease the coupling laser intensity.
Note that when the ratio of $\bar{\Omega}_1/
\bar{\Omega}_2$ is close to unity, one had better
directly deal with  the original formula (\ref{e17}),
rather than using its power series expansion.

In order to get some flavor of  the magnitude of the giant
nonlinearities derived above, we calculate
the nonlinear refractive-index coefficients
using the parameters ($I_2=40$ {\rm mW}/{\rm c$m^2$},
$\Delta_1=1.3\times 10^6$ rad/s, $\Delta_2=0$,  and
$\lambda_1=589$ nm) in the ultraslow light
experiment reported in Ref. \cite{hau}, in
which a light pulse speed $17$ m/s was
observed. We estimate that under these
conditions, $n^{(s)}_2=1.9\times 10^{-7}$ m$^2$/V$^2$,
 $n^{(s)}_4=-3.8\times 10^{-12}$ m$^4$/V$^4$, and
 $n^{(s)}_6=6.7\times 10^{-17}$ m$^6$/V$^6$. In terms of a common practical unit,
$n^{(s)}_2=0.36 {\rm cm}^2/{\rm W}$,
$n^{(s)}_4=-13.0  {\rm cm}^4/{\rm W}^2$,
and $n^{(s)}_6=4.5\times 10^2 {\rm cm}^6/{\rm W}^3$.
Hence, this value of  the Kerr nonlinearity is consistent with
 that indirectly measured
in Ref. \cite{hau}, almost $10^6$ times
greater than that measured in cold Cs atoms
\cite{hau,lam}, and $\sim 10^{12}$ times greater
than that measured in other materials
\cite{sal}. The fourth-order refractive-index
coefficient $n^{(s)}_4$ is $\sim 10^{22}$ times
greater than that measured in other
materials \cite{sal} .

The second nonlinear constants $n_4$, known as the $\chi^{(5)}$
nonlinear index of refraction,  is also an important parameter
in contemporary nonlinear optics. Theoretical investigations show that
in order to obtain spatial solitary waves a certain relation between
 $n_2$ and $n_4$ must be satisfied. Specially, Wright and his coworkers
 \cite{wri} indicated that  the ratio between the
second- and the fourth-order refractive-index
coefficients is an essential parameter
to obtain stable  spatial solitary waves.  The lower
the ratio $n^{(s)}_2/n^{(s)}_4$, the lower the required
power for stable beam propagation. For the
atomic medium with EIT, we have $n^{(s)}_2/n^{(s)}_4
=-I_2/(3\epsilon_0c)$.
With the parameters in the experiment\cite{hau},
we estimate $|n^{(s)}_2/n^{(s)}_4|\sim 10^{-2} {\rm W}/{\rm cm}^2$.
This ratio is small compared to most other nonlinear media
\cite{sal} by almost 11 orders of magnitude.
From Eq. (\ref{e30}) we can also obtain the ratio
between the fourth- and the sixth-order
refractive-index coefficients
$n^{(s)}_4/n^{(s)}_6=-3I_2/8(\epsilon_0c)$,
which is of the same order as that of $n^{(s)}_2/n^{(s)}_4$.

%%%%%%%%%%%%%%%%%%%%%%%%%%%%%%%%%%%%%%%%%%%%%

\section{Response to nonclassical fields}
%%%%%%%%%%%%%%%%%%%%%%%%%%%%%%%%%%%%%%%%%%%%%

We now discuss  the influence of
non-classical light on atomic polarization.
Allowing such a study is one of the advantages
of our fully quantum treatment of both the
coupling and the probe laser fields. We assume
that initially the atom is in the ground
state, two laser fields in an arbitrary state
$\sum^{\infty}_{n_1,n_2=0}C_{n_1}C_{n_2}
|n_1,n_2\rangle$, and the coupling and probe
fields are adiabatically applied. The
"steady"  state of the system is found to be
\begin{equation}
\label{e58}
|\Psi(t)\rangle=\sum^{\infty}_{n_1,n_2=0}
C_{n_1}C_{n_2} \exp\left [-ie^{(0)}_{n_1,n_2}t\right ]
|\phi^{(0)}_{n_1,n_2}\rangle.
\end{equation}

The Fourier component of the optical coherence
at probe-laser frequency is then given by
\begin{equation}
\label{e59}
\rho^A_{21}(\omega_1)=\sum^{\infty}_{n_1,n_2=0}
C_{n_1+1}C^*_{n_1}|C_{n_2}|^2
a_0(n_1,n_2)b_0(n_1,n_2).
\end{equation}
which indicates that the atomic polarization depends upon the
single-photon coherence of the initial probe field, i.e.,
$C_{n_1+1}C^*_{n_1}$. Hence, the atomic polarization is zero, if
the probe filed is initially in a state of no single-photon
coherence.

This point can be understood by the following
argument. The polarization is produced by the
electric field. When the probe field is initially
in a state of no single-photon coherence, the
mean value of the electric field in the state
vanishes, therefore the polarization of the
medium is zero. When the coupling and probe
lasers are in the coherent state
$|\alpha,\beta\rangle$, Eq. (\ref{e59}) reduces
to Eq. (\ref{e15}) in the large-$n$ approximation.
Obviously, the large-$n$ approximation should be given up to
investigate  on the response of the EIT medium to
nonclassical laser fields.

\section{ Concluding remarks}
In conclusion we have studied the linear and nonlinear optical
properties of EIT medium  interacting with two quantized laser
fields for the adiabatic EIT case in terms of a time-independent
approach in which both probe and coupling lasers are included as
parts of the dynamical system under our consideration, and treated
on the same footing. It has been shown that the fully-quantized
treatment for both probe and coupling lasers can not only
describes the ultraslow light experiments \cite{hau} very well,
but also sheds new insight for the response of EIT media to
nonclassical laser fields. In fact, we have found that very good
agreement with experimental results for slow group velocities and
the nonlinear refractive index of the probe laser observed in
experiments \cite{hau}. We have shown that EIT medium exhibits
normal dispersion. We have investigated the group velocities of
both probe and coupling lasers in adiabatic EIT media. It has been
found that the group velocities of both probe and coupling lasers
are reduced. It should be pointed out that there are conditions
for achieving slow group velocities of both probe and coupling
lasers. In the perturbation regime, the
weak-probe-strong-coupling-laser configuration leads to the fact
that only the probe laser have significantly reduced group
velocity. When using a strong-probe-strong-coupling-laser
configuration, one transfers appreciable population to the excited
state $\vert 3\rangle$ at the early stage of the pulse. This is
the familiar coherent population transfer stage. After this stage,
both probe and coupling lasers take dual role: they both work
partly as a probe and partly as a control laser. The consequence
is that both will experience slow down. We have studied
refractive-index changes of both probe and coupling lasers in the
vicinity of the resonance. We have also calculated nonlinear
susceptibilities and nonlinear refractive-index coefficients in a
completely analytical form. We have indicated that EIT medium
exhibits giant resonantly enhanced nonlinearities. We have
discussed the response of the EIT medium to nonclassical light
fields, and indicated that the polarization vanishes when the
probe laser is initially in a nonclassical state of no
single-photon coherence.

Finally, it should be remarked that in our treatment of EIT, which
incorporates both probe and coupling lasers as part of the
dynamical system, the decay parameters of various levels are
ignored. This is a good approximation, since EIT is insensitive to
any possible decay of the top level $|2\rangle$: During the
adiabatic preparation the population of the level $|2\rangle$ in
the dressed state $\phi^{(0)}$ remains negligibly small; see eqs.
(\ref{e6}) and (\ref{e8}). Even if we add by hand an imaginary
part to the energy of the level $|2\rangle$, it will {\it not}
enter the energy eigenvalue for the drssed state $\phi^{(0)}$.
Moreover, the effects of other decay parameters are expected to
depend only on their ratios to $\bar{\Omega}_1$ or
$\bar{\Omega}_2$, which normally are too small to dramatically
change our results.

\acknowledgments

This work was supported in part by the US NSF under Grant No.
PHY-9970701, and a Seed Grant from the University of Utah. L.M.K.
also acknowledges the National Fundamental Research Program
(2001CB309310), the China NSF under Grant Nos. 90203018 and
10075018, the State Education Ministry of China and the
Educational Committee of Hunan Province.


\begin{references}
\bibitem{har}  See, e.g., S.E. Harris, Phys. Today {\bf 50}, No. 7, 36 (1997;
                M.D. Lukin and A. Imamo\v{g}lu, Nature {\bf 413}, 273 (2001).
\bibitem{aga} G.S. Agarwal and W. Harshawardhan,
                Phys. Rev. Lett. {\bf 77}, 1039 (1996);
               M.O. Scully and M. Fleischhauer,
                {\it ibid.} {\bf 69}, 1360 (1992);
                M.D. Lukin and A. Imamo\v{g}lu,
                 {\it ibid.} {\bf 84}, 1419 (2000);
               M. Fleischhauer and  M.D. Lukin,  A.B. Matsko, and  M.O. Scully,
                {\it ibid.} {\bf 84}, 3558 (2000);
               B.S. Ham and P.R. Hemmer,
                {\it ibid.} {\bf 84}, 4080 (2000);
               A.B. Matsko, Y.V. Rostovtsev, H.Z. Cummins, and M.O. Scully,
                {\it ibid.} {\bf 84}, 5752 (2000);
                 H. Schmidt and R.J. Ram,
               Appl. Phys. Lett. {\bf 76}, 3173 (2000).
\bibitem{ari}  E. Arimondo,  Prog. Opt. {\bf 35}, 2570 (1996);
               S. Alam, {\it Laser without Inversion and Electromagnetically
               Induced Transparency} (SPIE-The international Society for
               Optical Engineering, Bellingham, Washington, 1999).
\bibitem{scu}  M.O. Scully and M.S. Zubairy, {\it Quantum Optics}
               (Cambridge University Press, England, 1997) p.220-244.
\bibitem{fie}  J.E. Field, K.H. Hahn, and S.E. Harris,
               Phys. Rev. Lett. {\bf 67}, 3062 (1991);
               K.-J. Boller,  A. Imamo\v{g}lu,  and S.E. Harris,
               {\it ibid.} {\bf 66}, 2593 (1991).
               L. Deng, M. Kozuma, E.W. Hagley, and M.G. Payne,
               {\it ibid.} {\bf 88}, 143902 (2002);
               L. Deng, M.G. Payne, and W.R. Garrett,
               Phys. Rev. A {\bf 64}, 023807 (2001).
\bibitem{hak}  K. Hakuta, L. Marmet, and B.P. Stoicheff,
               Phys. Rev. Lett. {\bf 66}, 596 (1991);
               G.Z. Zhang, K. Hakuta, and B.P. Stoicheff,
               {\it ibid.} {\bf 71}, 3099 (1993).
\bibitem{gea}  J. Gea-Banacloche, Y.Q. Li, S.Z. Jin, and Min Xiao,
               Phys. Rev. A {\bf 51}, 576 (1995);
               Y.Q. Li and Min Xiao, {\it ibid.} A {\bf 51}, R2703 (1995);
               Min Xiao, Y.Q. Li, S.Z. Jin, and J. Gea-Banacloche,
               Phys. Rev. Lett. {\bf 74}, 666 (1995);
               J. Qi, F.C. Spano, T. Kirova, A. Lazoudis, J. Magnes, L. Li, L.M. Narducci,
               R.W. Field, and A.M. Lyyra, {\it ibid.} {\bf 88}, 173003 (2002).
\bibitem{hau}  L.V. Hau, S.E. Harris, Z. Dutton, and C.H. Behroozi,
               Nature (London) {\bf 397}, 594  (1999).
\bibitem{ino}  S. Inouye, R.F. L\"{o}w, S. Gupta, T. Pfau, A. G\"{o}rlitz,
               T.L. Gustavson, D.E. Pritchard, and W. Ketterle,
                Phys. Rev. Lett. {\bf 85}, 4225 (2000).
\bibitem{kas}  M.M. Kash, V.A. Sautenkov, A.S. Zibrov,
               L. Hollberg, G.R. Welch, M.D. Lukin, Y. Rostovtsev,
              E.S. Fry, and M.O. Scully, Phys. Rev. Lett.
               {\bf 82}, 5229 (1999).
\bibitem{bud}  D. Budker, D.F. Kimball, S.M. Rochester, and V.V. Yashchuk,
               Phys. Rev. Lett.
               {\bf 83}, 1767 (1999).
\bibitem{tur}  A.V. Turukhin, V.S. Sudarshanam, M.S. Shahriar, J.A. Musser,
               B.S. Ham, and P.R. Hemmer, Phys. Rev. Lett. {\bf 88}, 023602 (2001).
\bibitem{hy}   S.E. Harris and Y. Yamamoto,
               Phys. Rev. Lett. {\bf 81}, 3611 (1998).
\bibitem{hh}  S.E. Harris and L.V. Hau, Phys. Rev. Lett.
                {\bf 82}, 4611 (1999).
\bibitem{sch}    H. Schmidt and A. Imamo\v{g}lu, Opt. Lett. {\bf 21},
                1936 (1996).
\bibitem{liu}  C. Liu, Z. Dutton, C.H. Behroozi, and  L.V. Hau,
               Nature (London) {\bf 409}, 490  (2001).
\bibitem{phi}  D.F. Phillips, A. Fleischhauer, A. Mair, R.L. Walsworth,
               and M.D. Lukin,  Phys. Rev. Lett. 86, 783 (2001).
\bibitem{nil}  M.A. Nielsen and I.L. Chuang, {\it Quantum Computation and
               Quantum Information}
               (Cambridge University Press, England, 2000).
\bibitem{fle}  M. Fleischhauer and M.D. Lukin,
               Phys. Rev. Lett. {\bf 84}, 5094 (2000).
\bibitem{kua}  Le-Man Kuang, Guang-Hong Chen,  and Yong-Shi Wu,
               quant-ph/0103152.
 \bibitem{mor} G. Morigi, J. Eschner, and C.H. Keitel,
               Phys. Rev. Lett. {\bf 85}, 4458 (2000);
               C.F. Roos, D. Leibfried, A. Mundt, F. Schmidt-Kaler,
               J. Eschner, and R. Blatt,  {\it ibid.}  {\bf 85}, 5547 (2000).
\bibitem{harr}  S.E. Harris, J.E. Field, and A. Imamo\v{g}lu,
                 Phys. Rev. Lett. {\bf 64}, 1107 (1990);
                K.H. Hahn, D.A. King, and S.E. Harris,  {\it ibid.}
                 {\bf 65}, 2777 (1990).
\bibitem{gell} M. Gell-Mann and F.E. Low, Phys. Rev. {\bf 84}, 350 (1951);
            M. Gell-Mann and M.L. Goldberger, Phys. Rev. {\bf 91}, 398 (1953).
\bibitem{harri}  S.E. Harris, J.E. Field, and A. Kasapi,
                 Phys. Rev. A {\bf 46}, R29 (1992).
\bibitem{flescu} M. Fleischhauer and    M.O. Scully,
                 Phys. Rev. A {\bf 49}, 1973 (1994).
\bibitem{rat}  U. Rathe, M. Fleischhauer, S.Y. Zhu, T.W. H\"{a}nsch,
               and    M.O. Scully , Phys. Rev. A {\bf 47}, 4994 (1993);
               G.P. Agrawal, {\it Nonlinear Fiber Optics},
               (Academic, San Diego, Calif., 1989) p.16.
\bibitem{lam}  A. Lambrecht, J.M. Courty, S. Reynaud, and E. Giacobino,
               Appl. Phys. B {\bf 60}, 129 (1995).
\bibitem{sal}  S. Saltiel, S. Tanev, and A.D. Boardman,
               Opt. Lett. {\bf 22}, 148 (1997).
\bibitem{wri}  E.M. Wright, B.L. Lawrence, W. Torruellas, and G. Stegeman,
               Opt. Lett. {\bf 20}, 2481 (1995).
\end{references}
\end{document}